\documentclass[a4paper]{jpconf}
\usepackage{graphicx}
\newcommand{\psim}{\lower.5ex\hbox{$\; \buildrel \propto \over\sim \;$}}
\newcommand{\lesssim}{\lower.5ex\hbox{$\; \buildrel < \over\sim \;$}}
\newcommand{\gtrsim}{\lower.5ex\hbox{$\; \buildrel > \over\sim \;$}}
\newcommand{\g}{\gamma}

\begin{document}
\title{ The obscured $\gamma$-ray and UHECR universe
}

\author{Charles D.\ Dermer}

\address{Code 7653, Naval Research Laboratory, 4555 Overlook Ave., SW, 
Washington, DC 20375-5352, USA}

\ead{dermer@ssd5.nrl.navy.mil}

\begin{abstract}
Auger results on clustering of $\gtrsim 60$ EeV 
ultra-high energy cosmic ray
(UHECR) ions and the interpretation of
the $\gamma$-ray 
spectra of TeV blazars
are connected by effects from the
extragalactic background light (EBL).
The EBL acts as an 
obscuring medium for $\gamma$ rays and a reprocessing 
medium for UHECR ions and protons, causing the GZK cutoff. The 
study of the physics underlying the 
coincidence between the GZK energy  and the clustering energy
of UHECR ions favors a composition of $\gtrsim 60$ EeV UHECRs
in CNO group nucleons. This has interesting implications
for the sources of UHECRs. We also comment on the Auger analysis.
\end{abstract}

\section{Introduction}

The recent reports \cite{aug07,aug08} from the Pierre Auger Observatory 
collaboration  
of high-significance
clustering of UHECRs along the supergalactic plane has opened 
multi-messenger astronomy. 
This advance is closely connected to progress in 
$\gamma$-ray and neutrino astronomy. 
Here we focus on new techniques to unravel the problem 
of UHECR origin in light of the Auger results. 
This topic is developed in more detail in my M\'erida ICRC contribution \cite{der07}
and in my book with Govind Menon \cite{dg07}.

\section{Charged Particle Astronomy}

We begin by reviewing the recent major results from Auger, including
the discovery of the clustering of the arrival directions
of UHECRs along the supergalactic plane \cite{aug07}. This discovery was anticipated
in the study by Stanev et al.\ \cite{sta95}, and we return  
to this paper in the discussion.

\subsection{Pierre Auger Observatory}

Auger is $\sim 3000$ km$^2$ in area, about the size 
of Rhode Island. It consists of 1600 shower detectors 
spaced 1.5 km apart, as well as four telescope enclosures
 each housing 6 telescopes to map Ni fluorescence. 
The directional point spread function of an UHECR is 
typically $\lesssim 1^\circ$. The original analysis
published in {\it Science} \cite{aug07} treated 
a  data set from 1 Jan 2004 to 31 August 2007 consisting
of  81 events with total energy $E > 40$ EeV. 

\subsection{Clustering Result}

A probability statistic $P$ corrected for exposure 
is constructed from the 
nearest-neighbor angular separation $\psi$ between
the arrival direction of an UHECR with energy $E$ and
the direction to AGN in the V\'eron-Cetty \& V\'eron (VCV) catalog \cite{vcv06}.
This catalog has 694 active galaxies with $z \leq 0.024$ or distance
$d \lesssim 100$ Mpc (the total catalog has over $10^5$ entries)
but is incomplete, especially
around the plane of our Galaxy, and is also
biased by different sample selection techniques.

In the Auger analysis \cite{aug07}, 
$P$ was minimized for $\psi =3.1^\circ$, threshold energy 
$E_{thr} = 56$ EeV, 
and $z_{max} = 0.018$ ($d_{max} = 75$ Mpc), containing 27 events
(the two highest energy events were 90 and 148 EeV).
Twelve events correlate within 3.1$^\circ$ of
 the selected $d<75$ Mpc AGNs, and another three within the vicinity
of one of these nearby AGN. By comparison,
3.2 events would be expected for an isotropic UHECR flux, 
which it clearly is not. Lack of correlation of $\gtrsim 60$ EeV 
UHECRs with 
the Galactic plane rules out Galactic
(pulsar, stellar mass black hole) and Galactic halo (including
halo dark matter annihilation) models, leaving extragalactic
models viable, at least those that follow the matter distribution
traced by the supergalactic
plane like the nearby AGN in the VCV catalog.

The recent detailed Auger analysis \cite{aug08} confirms these 
results, furthermore pointing out
that the distance cutoff could range from $\sim 50$ 
-- 100 Mpc and that $\psi \lesssim 6^\circ$.
If one accepts the validity of these results, then
{\it UHECRs with $E\gtrsim 60$ EeV originate from AGNs 
within
$75(\pm 25)$ Mpc and are deflected in their travels by at 
most $\sim 3^\circ$ -- $6^\circ$} by either the Galactic
magnetic field \cite{aug08} or the intergalactic magnetic field
\cite{der07}.

The lack of a quasi-isotropic background UHECR
flux above $E> E_{cl}\cong 60$ EeV is most simply explained if 
more distant UHECRs fail to arrive here at Earth. 
Because we know that there are sources of UHECRs with energy 
up to at least 300 EeV \cite{bir95}, 
it follows that unless we were in a unusual over-density of
UHECR sources within $\sim 100$ Mpc and an under-density within $\sim 100$
-- 300 Mpc or more, then energy losses of UHECRs accelerated by and injected from 
more sources more distant than $\approx 50$ -- 100 Mpc are the
cause of the Auger anisotropy. 

The clustering energy $E_{cl} = 60$ EeV stands out in this analysis, 
separating $E\lesssim E_{cl}$ UHECRs formed mainly in the large scale 
quasi-homogeneous, isotropic universe at distance scales $d \gtrsim 50$ -- $100$ Mpc
from the UHECRs formed in the clumpy structured
universe at local, $d\lesssim 100$ Mpc scales.

\subsection{GZK Cutoff}

\begin{figure}[h]
\includegraphics[width=18pc]{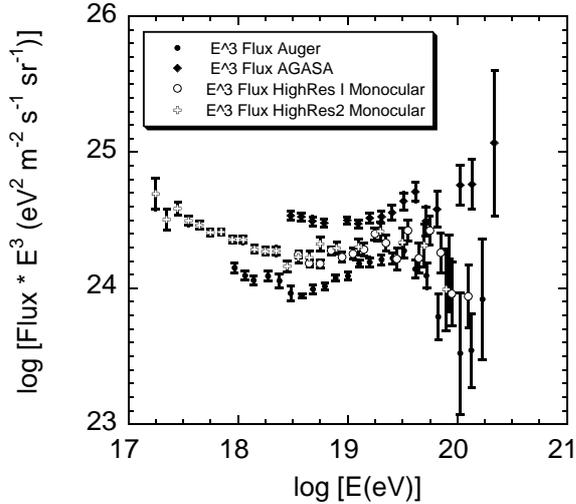}\hspace{2pc}%
\begin{minipage}[b]{10pc}\caption{\label{f1}UHECR spectrum, plotted 
in the form of $E^3 I$, where $I$ is the number intensity, for 
AGASA (filled diamonds), HiRes (open symbols), and 
Auger (filled circles).}
\end{minipage}
\end{figure}

A high-significance steepening in the UHECR spectrum at the Greisen-Zatsepin-Kuzmin
(GZK) \cite{gre66,zk66}
energy $E_{\rm GZK}\cong
10^{19.6}$ eV $\cong 4\times 10^{19}$ eV was reported  at the 2007 M\'erida ICRC based
on observations taken with the Auger Observatory \cite{yam07} (see Figure \ref{f1}), 
and earlier in 2007 by the HiRes 
collaboration \cite{hires07} at $E_{\rm GZK}\cong
10^{19.8}$ eV $\cong 6\times 10^{19}$ eV.\footnote{The HiRes team also reports 
a significant dip structure at $E_{ dip}\cong
10^{18.6}$ eV.} This observation seems to confirm the prediction \cite{gre66,zk66}
that interactions of UHECRs with CMBR photons will cause a break in the UHECR 
spectral intensity near $10^{20}$ eV. The coincidence $E_{\rm GZK}\cong E_{cl}$ seems likely to 
originate from underlying physics.

\subsection{UHECR Composition}

The composition of UHECRs measured from $\approx 4.5\times 10^{17}$ eV
up to $4\times 10^{19}$ eV with Auger 
is neither dominant p nor Fe composition \cite{ung07}, and is not trending
toward a lighter composition at the upper energy range, contrary
to pre-Auger indications \cite{wat06}.
If the results of Ref.\ \cite{ung07} can be extrapolated to $\gtrsim 60$ EeV, 
then ions must be included in the analysis unless the UHECR composition 
abruptly shifts to proton-dominated at $E\gtrsim E_{cl}$, as 
would be the case in neutral beam models of UHECRs \cite{ad03}.

\section{GZK Energy}

The energy of the GZK cutoff is not precisely defined. Returning to the classic paper
by Greisen \cite{gre66} for inspiration, let us work through the estimates
in light of our new knowledge.

\subsection{GZK Energy: Analytic Analysis}

UHECR protons with energy $E_p=\gamma_p m_pc^2 = 
10^{20}E_{20}$ eV will undergo strong photopion losses when the threshold 
condition $\gamma_p\epsilon_{CMBR}\gtrsim m_\pi/m_e \cong 400$ is satisfied, where
$\epsilon_{CMBR} \cong 2.70\times 2.72 {\rm ~K}/ 5.9\times 10^9 {\rm ~K} \cong  10^{-9}$
is the dimensionless mean photon energy, in units of the electron rest mass, 
of the CMBR. Strong photopion losses due to interactions with the peak energy photons of the
CMB occur when $\gamma_p \gtrsim 4\times 10^{11}$.
The mean free path (MFP) for energy loss of 
a proton with energy $E_{20} \gtrsim 4$ is 
$\lambda_{\phi\pi} \cong (1/ n_{CMBR} \sigma_{\phi\pi})\cong {1/ (400 {\rm~cm}^{-3}\cdot 70\mu{\rm b})}
\cong 10 {\rm~Mpc}$. From a derivation using a step-function 
approximation for the product of photopion
cross section and inelasticity \cite{ad03}, the mean-free-path for photopion energy losses of UHECR protons 
with CMBR photons is given by \cite{der07}
\begin{equation}
\lambda_{\phi\pi}(E_{20}) \cong {13.7 \exp[4/E_{20}] \over [1 +
4/E_{20}]}\;{\rm Mpc}\;.
\label{eq1}
\end{equation}
The values of $\lambda_{\phi\pi}(E_{20})$ for $E_{20} = 0.6, 0.8, 1,$ and 2 
are $1400, 340, 150$, and 35 Mpc, respectively. If the 
UHECR with energies between 60 and 100 EeV were protons, then they 
could originate over a large
distance scale extending to 1 Gpc. This simple estimate
suggests that the the UHECRs cannot be protons because
the Auger events originate from within
100 Mpc whereas the energy loss MFPs of UHECR protons 
at these energies are much greater. 

Now we estimate the GZK MFPs of ions against photodisintegration on
CMBR photons. Note that by contrast with UHECR protons, photopion losses of UHECR nucleons 
are not effective until
$\gamma \langle \epsilon \rangle_{CMB} \gtrsim m_\pi/m_e$ or
$E_{20} \gtrsim 3A/(1+z)$, where $A$ is the nucleon mass number. The pion-forming 
$\Delta$ resonance peak energy is $\cong m_ec^2\epsilon_r \approx 300$
MeV, a much higher energy than the giant dipole resonance (GDR) at
$\approx 20$ MeV, so photodisintegration breakdown usually dominates
photopion losses for ions, especially for large $A$. 
The  effective photodisintegration 
energy-loss rate for ions on the CMBR results from both GDR excitation
with the emission
of one and two nucleons, and from multi-nucleon breakup processes. 
Because the reaction is a threshold process
and UHECRs with $E_{20} \ll 3A/(1+z)$ interact only with
the Wien portion of the CMBR radiation field, the GDR is more important
than multi-nucleon excitations for $E_{20} \sim 1$ UHECRs. 

In this asymptote, we use the 
$\delta$-function approximation for the 
photodisintegration cross
section of nucleus $A$, namely
\begin{equation}
\sigma_A(\epsilon_r ) \cong {\pi\over 2} 
\sigma_{0,A} \Delta_{GDR} \,\delta(\epsilon_r - \epsilon_{r,A})\;,
\label{sigdelta}
\end{equation}
\cite{kt93,anc07,wrm07},
where 
$$\sigma_{0,A} = 1.45A {\rm~mb}\;, \;\;\Delta_{GDR} \cong 15.6\;,\;
\epsilon_{r,A}\cong 83.5 A^{-0.21},$$
and $\epsilon_r =\gamma\epsilon(1-\mu)$ is the invariant dimensionless photon
energy in the ion's rest frame. 
In the $\delta$-function approximation for the GDR, eq.\ (\ref{sigdelta}),
the derived inverse of the 
effective energy loss time scale of an UHECR ion due to photodisintegration 
processes with photons of the CMBR is 
$$t_E^{-1}(E,A) = {2\pi^2 c \Delta_{GDR} k_A \epsilon_{r,A}\Theta\over
\gamma^2 A \lambda_{\rm C}^3} \;\ln\big([1-\exp(-w_{r,A})]^{-1}\big)$$
\begin{equation}
\cong 3.2\times 10^{-15}\;{k_A A^{1.79}(1+z)\over E_{20}^2} \;\ln\big[{1\over 1-\exp(-w_{r,A})}\big]\;{\rm~s}\;,
\label{tEEA}
\end{equation} 
where 
$$w_{r,A} \equiv {\epsilon_{r,A} \over 2\gamma \Theta} = {0.83 A^{0.79}\over
E_{20}(1+z)}\;.$$
The effective path length for photodisintegration in the $\delta$-function approximation 
for the GDR is therefore
\begin{equation}
\lambda_E = \lambda_E(E,A) = c t_E(E,A)\rightarrow 
{3.0E_{20}^2 \exp\big( 0.83A^{0.79}/[(1+z)E_{20}]\big)
\over k_A A^{1.79} (1+z)}\;\;{\rm Mpc~}
\label{lambdaE}
\end{equation}
in the limit $w_{r,A} \gg 1$, that is, $E_{20} \ll 5(A/10)^{0.79}$.

When a
single proton or neutron is ejected, then a fraction $A^{-1}$ of the
original energy $E$ is lost to the original nucleon, and for the ejection of
two nucleons, a fraction $ 2/A$ of energy is lost.  For multi-nucleon
injection, an average fractional energy loss $k_A/A$ is used \cite{psb76}, 
where $k_A = 1.2/A, 3.6/A,$ and $4.349/A$ for $A = 4, 10 \leq A \leq 22$, and 
$23 \leq A \leq 56$, respectively. The photodisintegration 
energy-loss MFPs have only a generalized
meaning in terms of the mean distance for an UHECR nucleon to 
be broken up into mostly lower energy protons and neutrons and a nucleon
with $A$ about half of the original nucleonic mass. Photodisintegration
of Fe, for example, leads to significant fraction of $A\lesssim 15$ nucleon 
secondaries \cite{psb76}.

Equating eq.\ (\ref{lambdaE}) with an energy-loss distance $\lambda = 100 \lambda_{100}$
for $E = 60 E_{60}$ EeV gives $\exp(1.38A^{0.79}/E_{60})/[k_A A^{1.79}]
 = 333 \lambda_{100} E_{60}^2$. It is easily verified that all nuclei with $A \gtrsim 14$
have energy-loss mean free paths at $E_{60} \approx 1$ much longer than 100 Mpc. On the basis of 
this analytic treatment, one might conclude
UHECRs could originate only from nucleons 
with $4\lesssim A \lesssim 14$.

\subsection{GZK Energy: Numerical Analysis}

\begin{figure}[h]
\includegraphics[width=24pc]{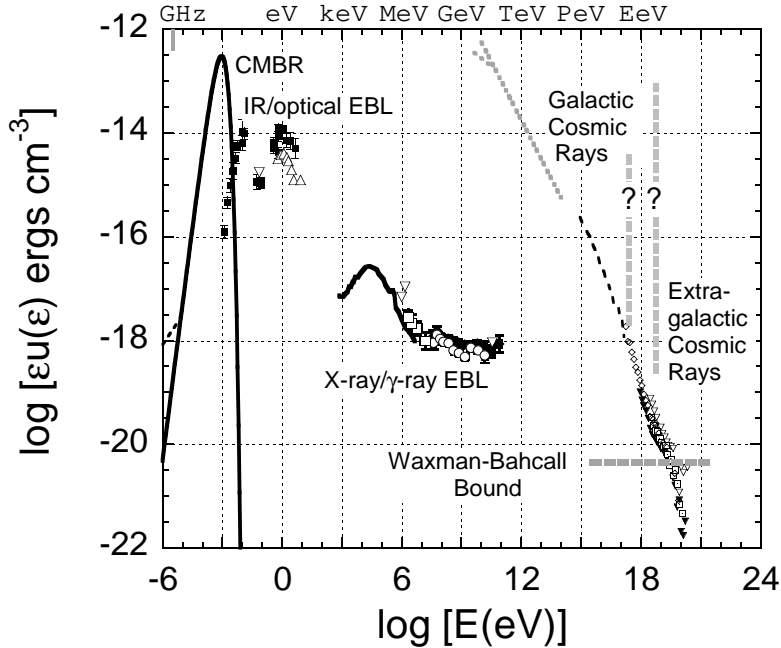}\hspace{2pc}%
\begin{minipage}[b]{10pc}\caption{\label{f2}Energy density in intergalactic space of various radiations,
including the CMBR, the IR and optical,  X-ray,
 $\gamma$ ray,  extragalactic cosmic ray, and the 
predicted maximum energy density of cosmogenic neutrinos. Also shown is the 
near-Earth energy
density of cosmic rays; the transition energy between the 
galactic and extragalactic component
is uncertain.}
\end{minipage}
\end{figure}

These results are not conclusive unless 
additional radiation fields, including IR and optical features
from stellar and reprocessed radiation,
and energy-loss processes of photopair production and universal expansion
are taken into account. Figure \ref{f2} shows the energy density 
of the EBL and cosmic rays, and the
predicted upper limit to the cosmogenic neutrino
energy density \cite{wb99}.  

\subsubsection{Extragalactic Background Light} 

\begin{figure}[h]
\begin{minipage}{14pc}
\includegraphics[width=18pc]{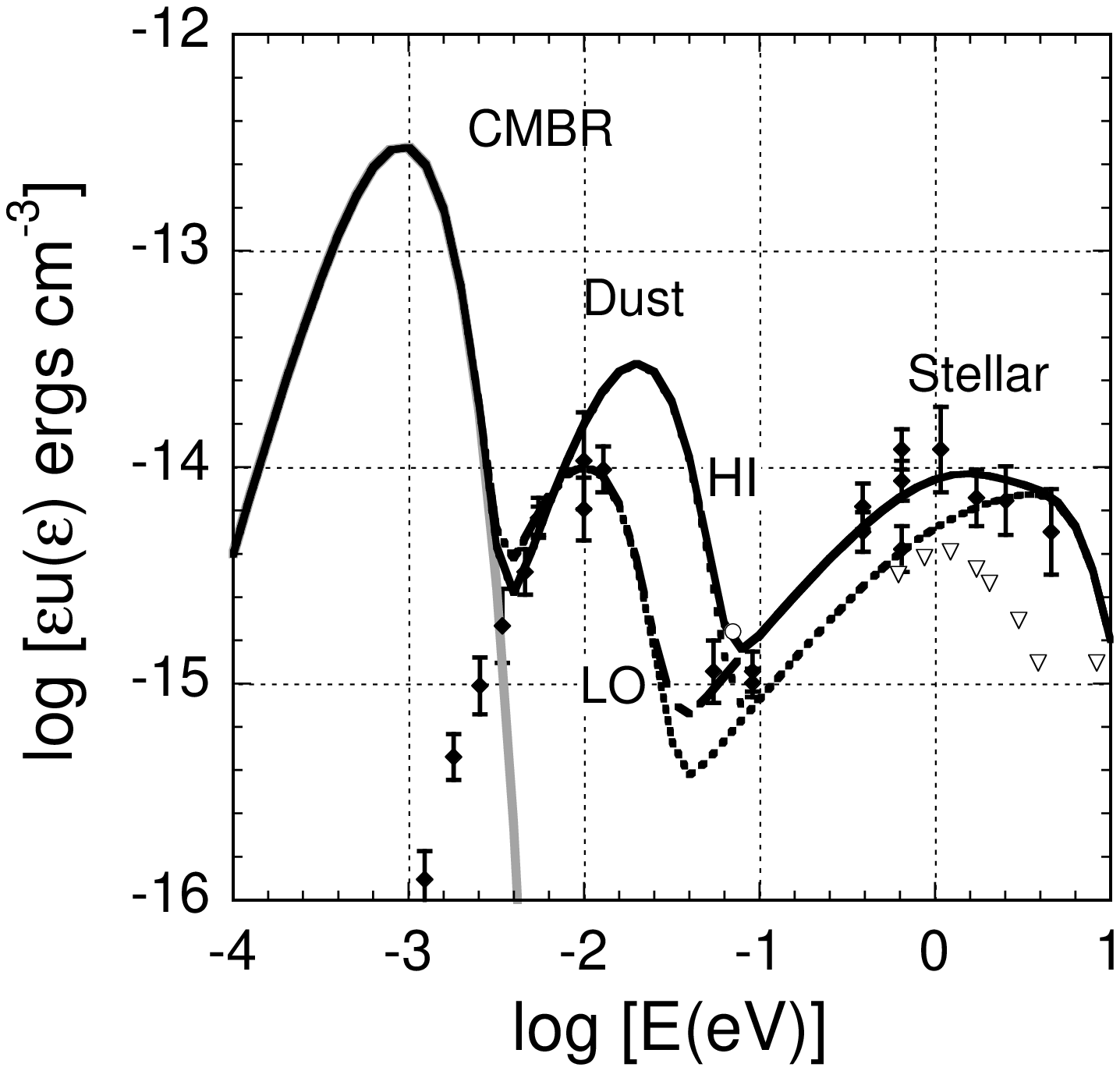}
\caption{\label{f3}Ranges of EBL used in analysis. Data from Ref.\ \cite{hd01}.
The HI (LO) EBL refers to a high-(low-) intensity dust component and a 
high-(low-) intensity
stellar component. }
\end{minipage}\hspace{6pc}%
\begin{minipage}{14pc}
\includegraphics[width=18pc]{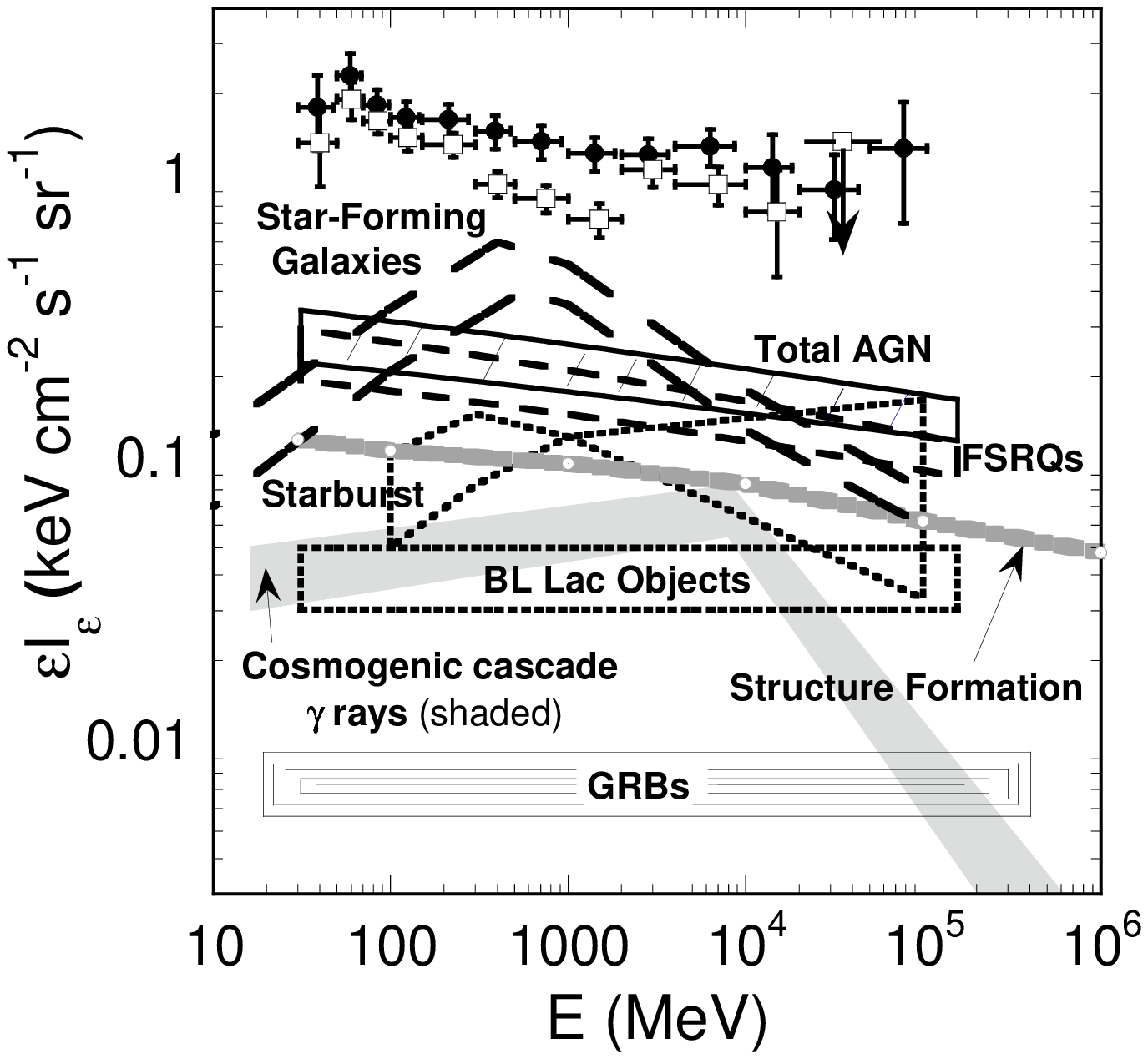}
\caption{\label{f4}Decomposition of the diffuse/unresolved $\gamma$-ray 
background measured with EGRET \cite{der07a}. Shaded region shows estimate for cascade $\gamma$-rays induced
by UHECRs formed by sources evolving as AGNs \cite{kss07}.}
\end{minipage} 
\end{figure}

The mean intensity of light in intergalactic space is 
referred to as the EBL. 
The IR and optical EBL is decomposed at low redshift ($z\lesssim 0.25$)
into a dust component and a stellar component as shown in Figure \ref{f3}. These fits are
assumed to span the likely range of the diffuse radiation fields between 
$\approx 1$ and a few hundred $\mu$m at low $z$. 
The modified blackbody spectral 
energy density is written in the form
\begin{equation}
\epsilon u(\epsilon ) = u_0\;{w^k\over \exp(w)-1}\;=\; m_ec^2 \epsilon^2 n_{ph}(\epsilon )\;, 
\label{nmodbb}
\end{equation}
where $w \equiv \epsilon/\Theta$. The stellar component
is decomposed into the sum of two modified blackbodies, with the higher-temperature stellar
component fixed. The HI EBL has the more intense dust and stellar fields, and the LO EBL has
the weaker dust and stellar fields. Two other EBLs are considered here: a high dust/low stellar field
and a low dust/high stellar field. Parameters of the CMBR and modified blackbodies are
given in Table 1. 

 \begin{table}
\begin{center}
\caption{Properties of the Dust and Two Stellar Components}
\begin{tabular}{lccc}
\hline
Component  & $T$(K) & $u_{0}$  & $k$ \\ 
 &  & ($10^{-14}$ ergs cm$^{-3}$) &  \\ 
\hline
\hline
 CMBR  &  2.72  &  6.38 & 4 \\
 Dust (Hi) & 62 & 0.819 & 3.8  \\
 Dust (Low) & 31 & 0.273 & 3.8  \\
 Star 1 (Hi)& 7100 & 1.1& 2.0  \\
 Star 1 (Lo)& 7100 & $0.55$ & 2.0  \\
Star 2 & 16,600 & 0.5 &  3.0 \\
\hline
\end{tabular}
\end{center}
\end{table}

The intensity of the diffuse and unresolved 
$\gamma$-ray EBL from 
EGRET observations and analysis \cite{sre98,smr04}
is shown in Figure \ref{f4}.
It is decomposed into
contributions from blazars, separated into FSRQs and
BL Lacs, and the superposed intensity from 
numerous faint sources where the $\gamma$ rays are formed primarily by
interactions of cosmic rays accelerated by supernovae in star-forming galaxies 
and by structure formation shocks in clusters of galaxies (reviewed
in \cite{der07a,kne07}). The particle and radiation fields are connected because
GZK effects of UHECRs with the EBL form cascade $\gamma$ rays that 
contribute a truly diffuse component to the $\gamma$-ray EBL \cite{kss07}.
The intensity of this component is
sensitive to the transition energy from galactic to extragalactic UHECRs
and the formation rate of UHECR sources through time.


\begin{figure}[h]
\begin{minipage}{14pc}
\includegraphics[width=14pc]{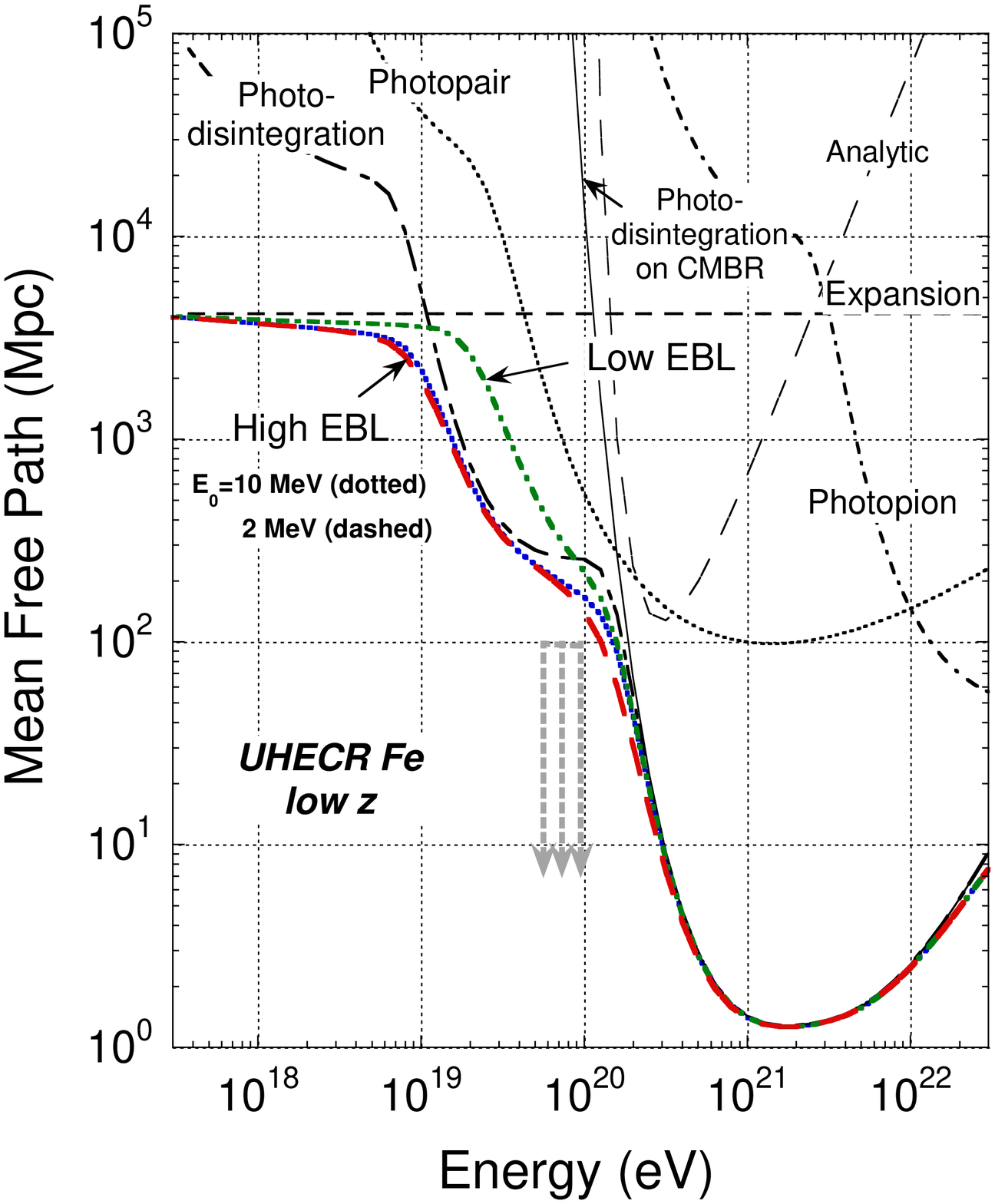}
\caption{\label{f5}Effective energy-loss mean-free-path of Fe in LO and HI EBLs, 
with the effective photodisintegration energy-loss MFP 
calculated according to the prescription of Ref.\ \cite{psb76}.  }
\end{minipage}\hspace{5pc}%
\begin{minipage}{14pc}
\includegraphics[width=14pc]{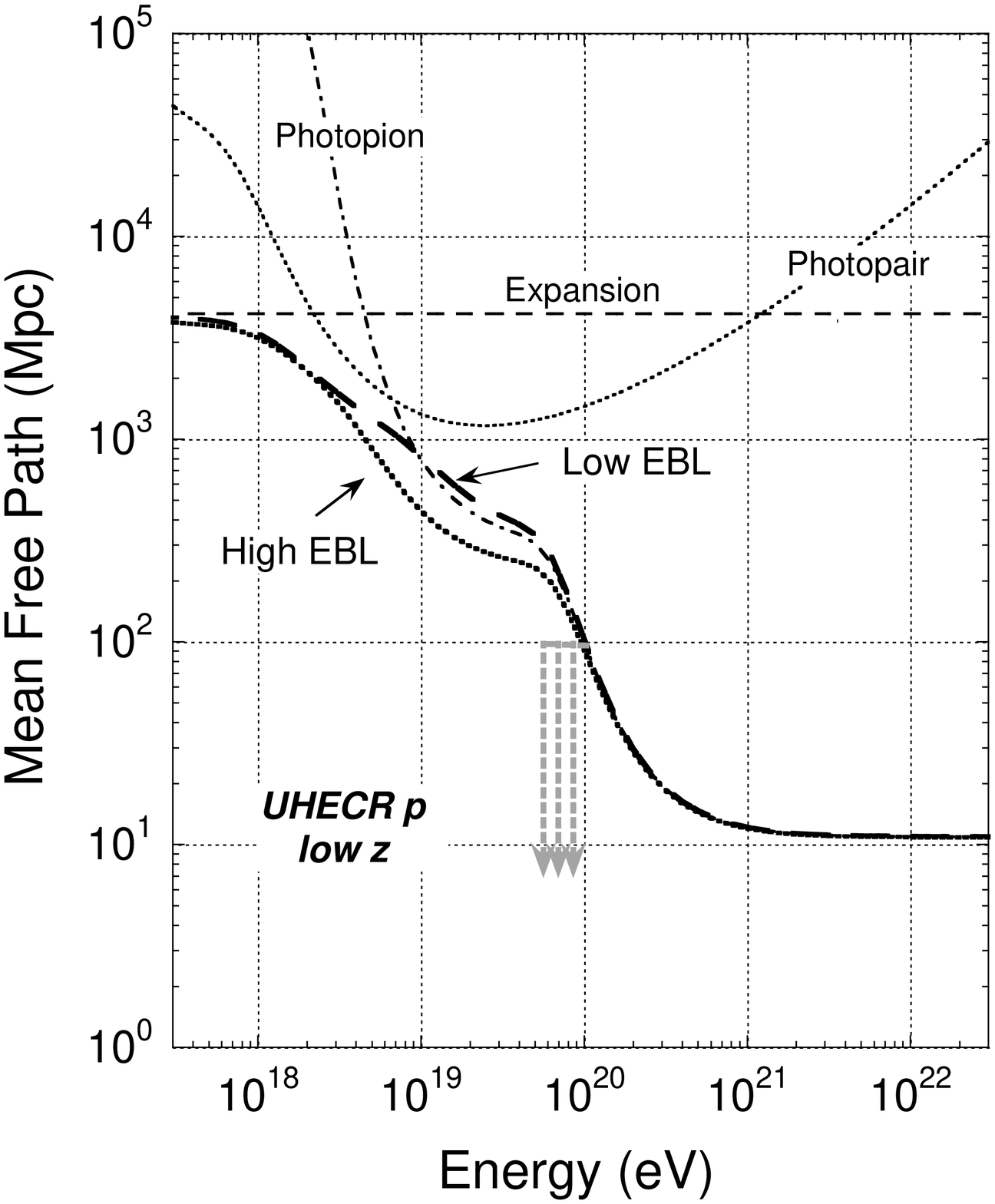}
\caption{\label{f6}Effective energy-loss mean-free-path of UHECR protons
in LO and HI EBLs shown in Fig.\ \ref{f3}.  }
\end{minipage} 
\end{figure}

The analytic result for the energy-loss
MFP of Fe with $A = 56$ on CMB photons is  given 
by eq.\ (\ref{lambdaE}) and labeled ``analytic"
in Figure \ref{f5}.
The accuracy is tolerable, but a more detailed calculation,
also shown in Figure \ref{f5}, 
is needed to draw more reliable conclusions. 
The numerical calculations for Fe treat
photopion and photopair losses 
with the EBL as well as from universal expansion (the Hubble
radius is $\approx 4160$ Mpc at the present epoch), with the
photodisintegration, photopair, and photopion components shown
separately for the HI EBL. The total energy loss
MFP is also shown for the LO EBL.  (See also \cite{hmr06,hst07}.)
The distances from which we should expect to detect 60 -- 100 EeV Fe
range from $\approx 200$ Mpc to 150 Mpc.

Fig.\ \ref{f6} shows the energy-loss MFP of UHECR protons in the 
LO and HI EBLs. The MFPs, which include photopion, 
photopair and expansion losses, range for the HI EBL from  
200 to 100 Mpc for energies ranging from 60 to 100 EeV, respectively
(similar to the values for UHECR Fe).
Unless there was a deficit of UHECR sources between $\approx 100$ -- 200 Mpc,
or an enhanced pathlength due to strong, $\gtrsim 10^{-10}$ G
intergalactic magnetic fields, it is difficult to account for unless 
UHECR p and Fe comprise only a small  
fraction 
of the $\gtrsim 60\times 10^{19}$ eV UHECRs detected with Auger.

How else to resolve the quandary that the UHECR proton and Fe MFPs are too long
to be commensurate with the Auger data, so that the UHECRs cannot
be protons or Fe? Possibly the Auger detector is not properly calibrated, 
though it is unlikely to be off by
more than $\approx 20$\% compared to the factor of $\approx 2$
needed. For instance, Berezinsky \cite{ber08}, 
using the dip energy seen most clearly in an $E^3 I$ plot
(see Fig.\ \ref{f1}) as a fiducial, and multiplying by an 
energy calibration factor 1.2, 0.75,
0.83, and 0.625 for Auger, AGASA, Akeno and Yakutsk
data, respectively, puts the various data sets in agreement with 
each other. Watson \cite{wat08} argues that there are $\approx 19$\% and
11\% errors in the Auger energy calibration
at $10^{17}$ and $10^{20}$ eV, respectively, so the correction factor for Auger
is $\lesssim 1.2$.
If this is true, then only a few percent of the $\gtrsim 6\times 10^{19}$ 
eV UHECRs could be protons.

Let us provisionally accept that the Auger clustering results and the
calculations presented here require that $\gtrsim 60$ EeV UHECRs are
depleted in p and Fe (a full Monte Carlo 
simulation is required  to remove remaining uncertainties).
Some models, in particular, 
the neutral beam model that Armen Atoyan and I developed (for blazars, see \cite{ad03}, 
and for GRBs, see \cite{da03}) were based on the acceleration of protons 
to  $\gtrsim 10^{20}/\Gamma$ eV energies in the shocked fluid frame,  where
$\Gamma$ is the flow Lorentz factor. 
Photopion processes make an escaping neutron and $\gamma$-ray beam. 
If protons do not make up a large fraction of the $\gtrsim 10^{19.6}$ eV UHECRs, 
then this model no longer has to contend with the 
difficulty of accelerating protons to such high energies. 
On the other hand, this or any other model must be able to accelerate
ions to ultra-high energies and get them from the accelerator into 
intergalactic space with the requisite emissivity.
 
\begin{figure}[h]
\begin{minipage}{14pc}
\includegraphics[width=14pc]{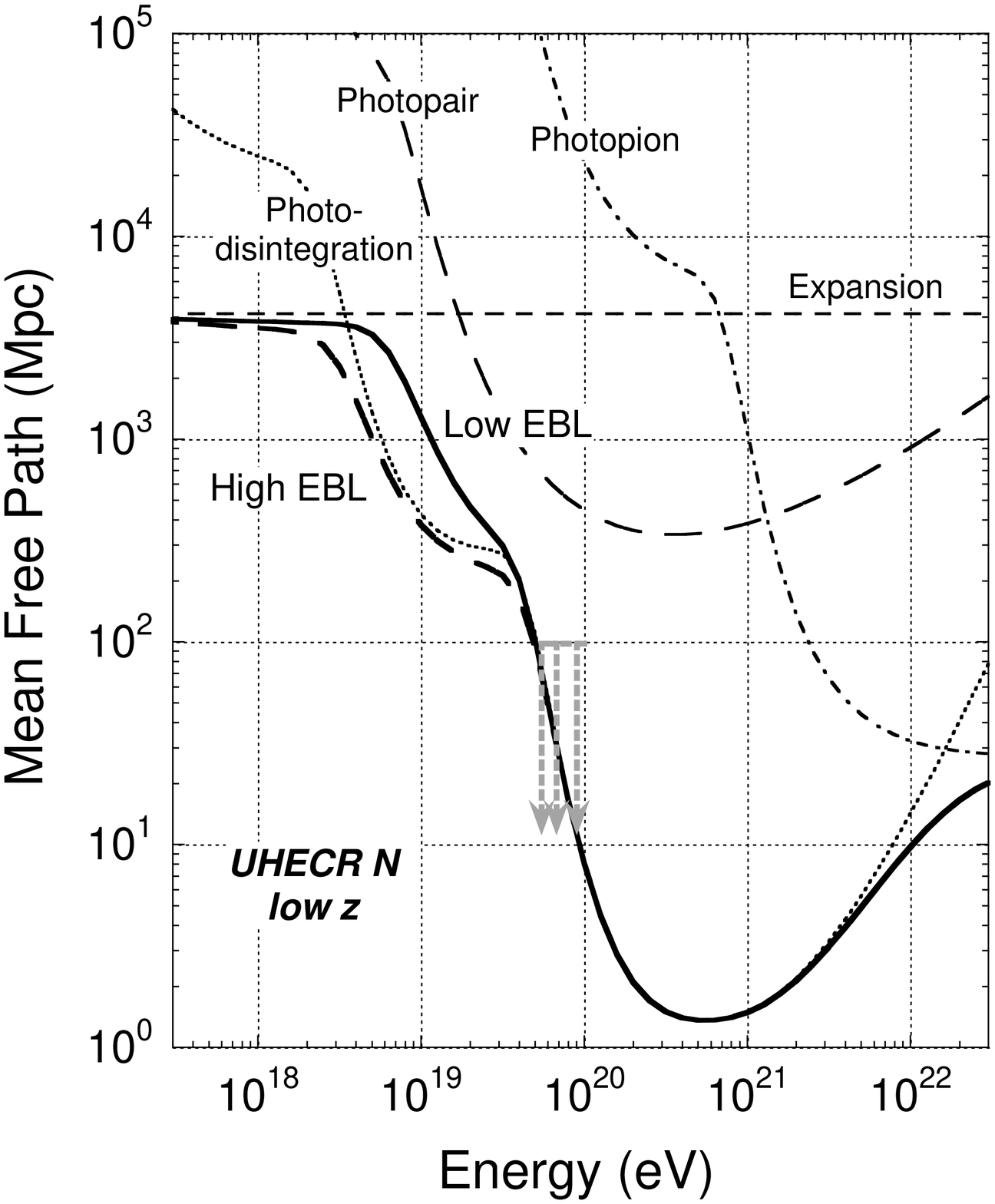}
\caption{\label{f7}As in Fig.\ (\ref{f5}), for UHECR N.  }
\end{minipage}\hspace{5pc}%
\begin{minipage}{14pc}
\includegraphics[width=14pc]{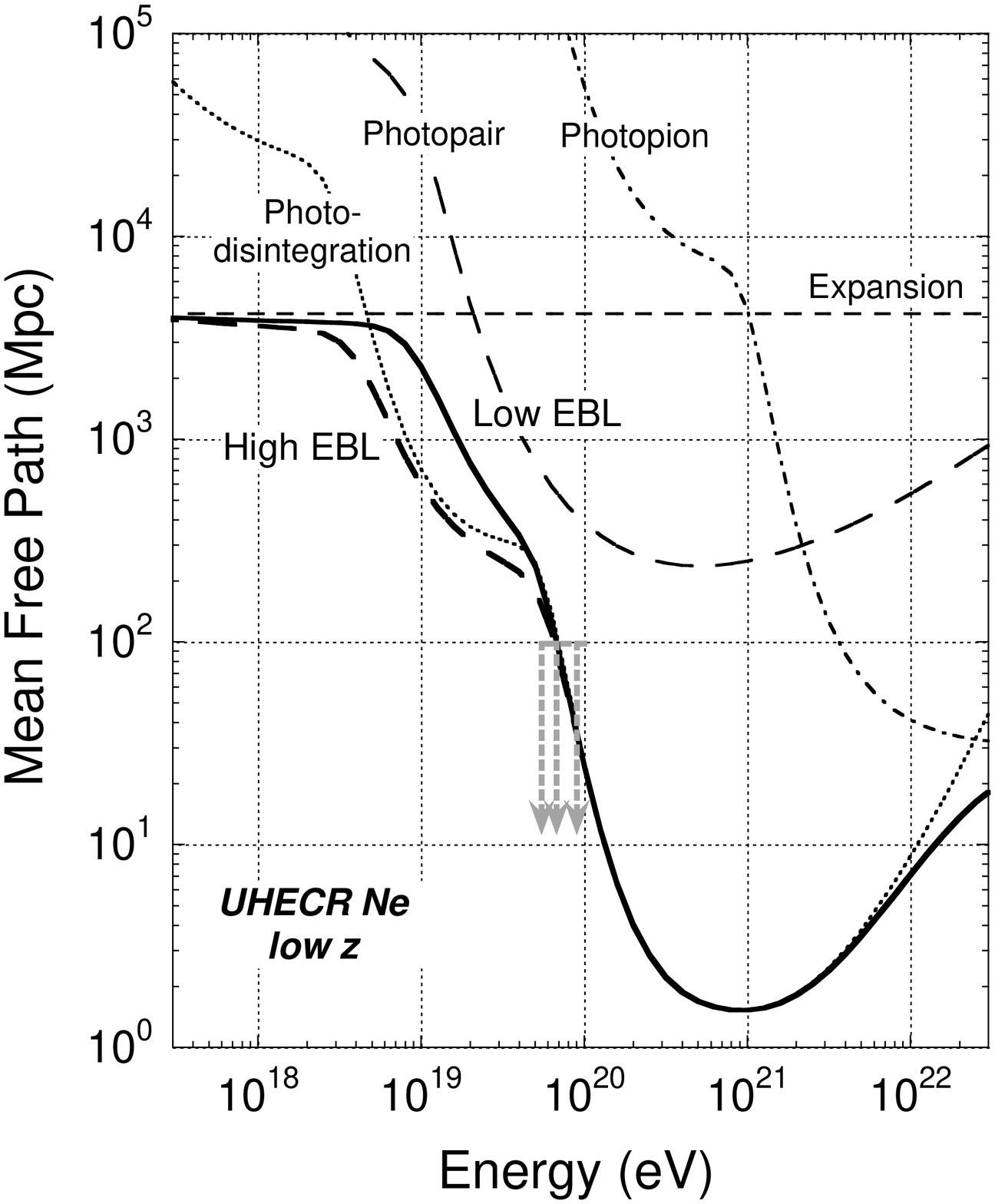}
\caption{\label{f8}As in Fig.\ (\ref{f5}), for UHECR Ne. }
\end{minipage} 
\end{figure}

Which ions? Again accepting the approximate integral meaning of an energy-loss MFP,
Figures \ref{f7} and \ref{f8} show that UHECR ions up to Ne ($A=20$) or at most
Mg ($A = 24$) would effectively 
lose their energy within $\approx 100$ Mpc. An abundance of UHECRs in light
ions gives the most economical 
explanation of the Auger results. If correct, 
then composition analyses of 60 -- 100 EeV UHECR data will give an average atomic mass 
number $\langle A \rangle \lesssim$ 24 ($\langle \ln A \rangle \lesssim 3.2$), and probably in the range $
\langle A \rangle
\lesssim 20$. This prediction can be compared with 
Fig.\ 10 in Watson \cite{wat08}, giving the measured energy dependence
of $\langle \ln A \rangle$ using the 
the QGSJETII-03 model.
According to our calculations, the anisotropy of $\cong 60$ EeV
UHECRs with $d_{max} \lesssim 50$ -- 100 Mpc 
could only originate from muclei with $2\lesssim \langle \ln A \rangle \lesssim 3$ (the lower limit is not firm, and
depends on source type).

\begin{figure}[h]
\begin{minipage}{14pc}
\includegraphics[width=18pc]{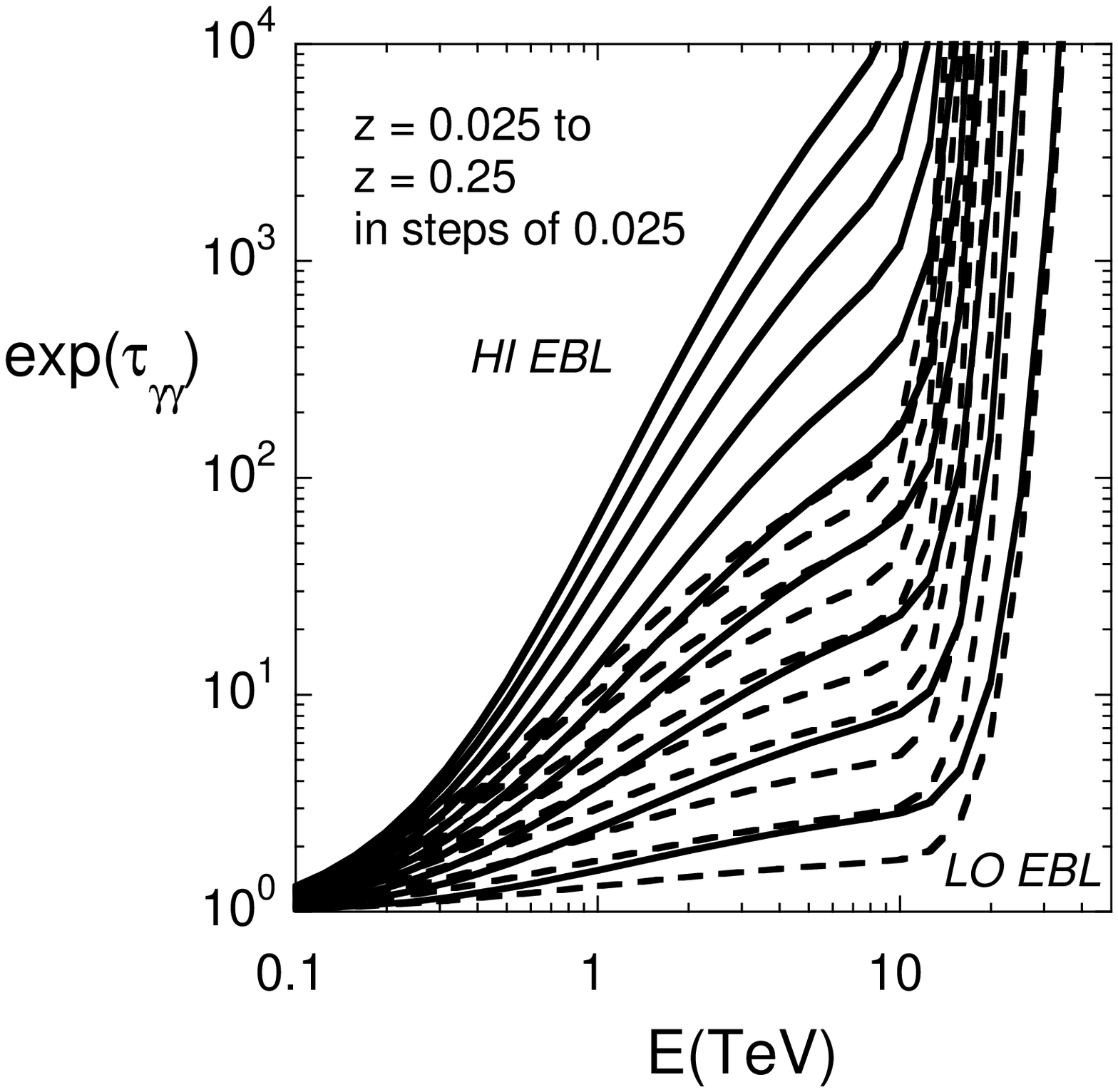}
\caption{\label{f9}The factor $\exp(\tau_{\gamma\gamma})$ at different
redshifts for the LO and 
HI EBLs shown in Fig.\ \ref{f3}, as a function of photon energy {\tt E}. }
\end{minipage}\hspace{5pc}%
\begin{minipage}{14pc}
\includegraphics[width=18pc]{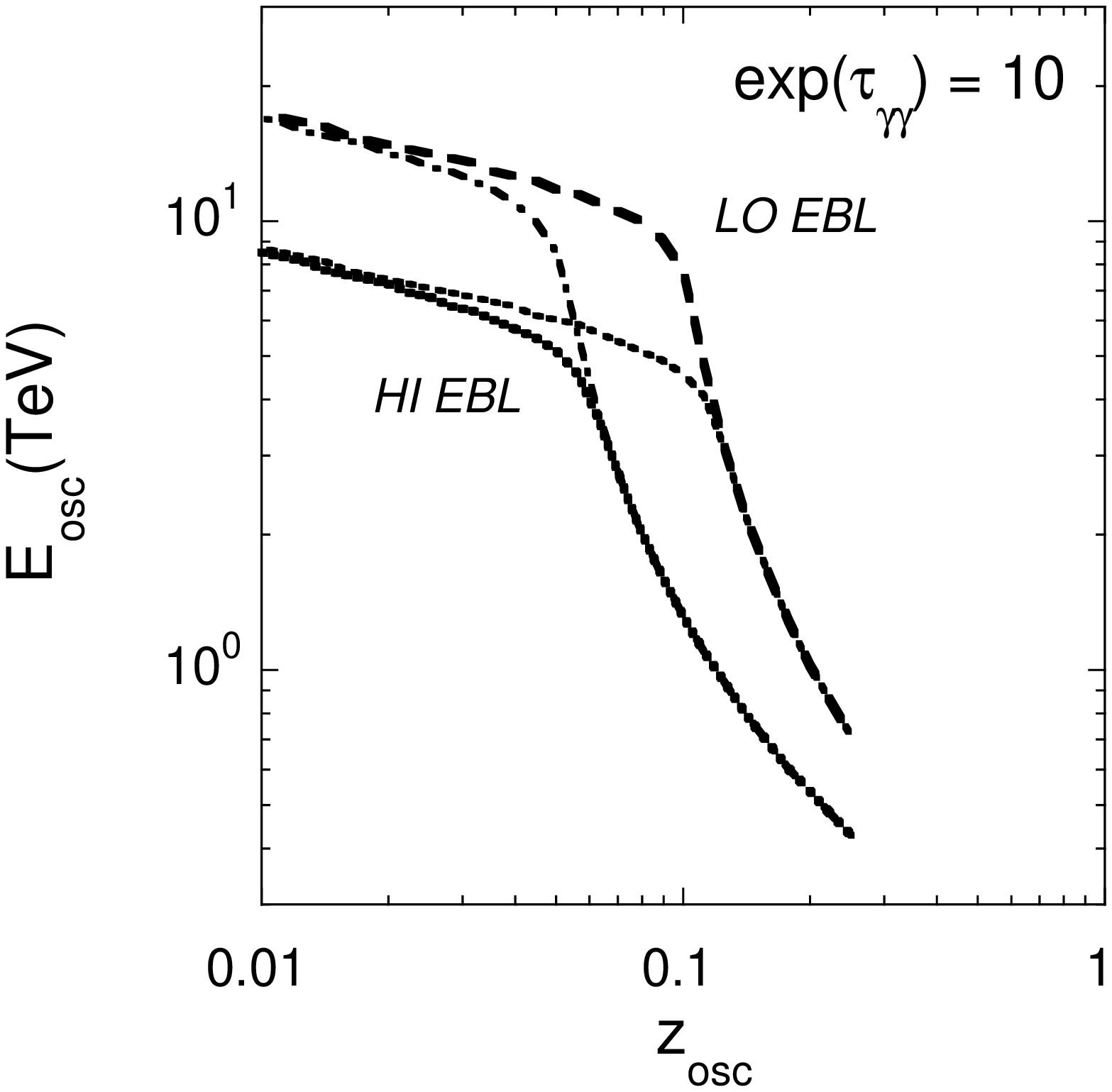}
\caption{\label{f10}Relation between the obscuration energy ${\tt E}_{osc}$ and 
obscuration redshift $z_{osc}$ defined by
 $\exp(\tau_{\gamma\gamma})= 10$
for various high and low dust and stellar EBLs.  }
\end{minipage} 
\end{figure}

Let us return to the question of the EBL.
The dust EBL primarily
attenuates $\gtrsim 4$ TeV photons and the stellar EBL primarily attenuates
$\lesssim 4$ TeV. The LO and HI EBLs can be used to calculate $\tau_{\g\g}$
up to redshifts $z \approx 0.25$ 
where cosmic evolutionary processes start to change the EBL energy density
compared to the present. 
The exponential attenuation factor $\exp(\tau_{\g\g})$
as a function of photon energy {\tt E} for different redshifts is plotted
in Figure \ref{f9}. We can make a cut defined by $\exp(\tau_{\g\g})
=10$ (or any other number, for that matter) to define a relationship 
between redshift and photon energy (the relation 
defined by $\tau_{\g\g}(z,{\tt E})=1$ is sometimes 
referred to as the ``Fazio-Stecker relation \cite{fs70}"; see also \cite{kne04})
Here we make the cut defined by $\exp(\tau_{\g\g})
=10$ to give the obscuration energy ${\tt E_{osc}}$ as a function of 
obscuration redshift ${\tt z_{osc}}$, shown in Fig.\ \ref{f10}.
This figure is not however very 
useful for comparison with observational data, 
because the highest measured energies of photons 
depend upon telescope sensitivity, and variable source spectral brightness. 
Nevertheless, the dust component forms an opaque 
screen to photons with ${\tt E}\gtrsim 7$ TeV and $\gtrsim 15$ TeV for the high and low
dust EBL at the 
redshift $z \approx 0.03$ of Mrk 421 and Mrk 501. Significant
detection of $\approx 8$ TeV from Mrk 421 \cite{fos07,yad07} and Mrk 501 \cite{sam98}, 
and of $\gtrsim 10$ TeV photons from 1ES 0229+200 \cite{aha07} ($z = 0.14$) 
suggests that the EBL intensity at $\sim 10$ -- $100~\mu$m
is smaller than given by the high dust EBL.

We  now ask why the sources of UHECRs should be enriched in 
medium mass (C -- Ne) nuclei. Refs.\
\cite{wrm07,mur08} discusses various scenarios for accelerating
UHECRs with significant ion
content in GRB environments, including capturing 
supernova shell material by a jetted relativistic wind, external
shock capture of circumburst material, and hypernovae \cite{wan07a}.
We favor the hypothesis
\cite{dm04} that an external shock is important
during the prompt and early
afterglow phases of a GRB. GRBs formed by core collapse of C or O 
Wolf-Rayet stellar progenitors would be surrounded by a wind
enriched in these light elements that could be captured
and accelerated to ultra-high
energies. This could be accomplished by 1$^{st}$- or 
2$^{nd}$-order Fermi processes
at the shock formed by the interaction of 
a relativistic wind with the surrounding medium.

For blazars and radio galaxies that 
propel intermittent winds to form internal
shocks, the composition of UHECRs would depend
on the makeup of the matter accreting onto 
supermassive black holes. If near-Solar metallicities
occur \cite{ag89}, then proton acceleration would have to 
be suppressed, perhaps by a gyroresonant mechanism, and 
the medium mass ions would have to be processed through the black hole
environment without breakup due to spallation or photodisintegration.

\section{Discussion}

The clustering result reported \cite{aug07}
by the Auger Collaboration is a major discovery, 
anticipated  
by Stanev, Biermann, Lloyd-Evans, Rachen, and Watson \cite{sta95} in 1995. 
In the earlier study, Northern hemispheric data from Haverah Park, AGASA, Volcano 
Ranch, and Yakutsk 
with 73 (27), 25 (7), 13 (5), and 32 (3) events above 20 (40) EeV, respectively, 
and with zenith angles smaller than
45$^\circ$ were analyzed. The average and rms angular distances of the arrival directions
of UHECRs from the supergalactic plane were calculated, after correcting for 
exposure. Compared to an isotropic source flux, a  statistical 
enhancement towards the supergalactic plane 
at the 2.5 -- 2.8 $\sigma$ level was found for arrival directions of events 
with $E> 40$ EeV. The supergalactic plane runs through the Virgo Cluster 
at $\approx 20$ Mpc, and contains an assortment 
of radio glaxies such as M87, Cen A and NGC 315, and the 
starburst galaxies M82 and NGC 253.
Stanev et al.\ \cite{sta95} argue that their analysis
 supports the hypothesis that radio galaxies are the sources of 
UHECRs accelerted, for exmple,  in the inner jets or at the knots or hot spots of radio galaxies. 
Supporting this contention 
is the analysis of Shaver and Pierre  \cite{sp89}
showing harder integral source count distributions 
of extragalactic radio sources at large spectral fluxes
toward the supergalactic plane in  the  
408 MHz $1$ Jy  southern hemisphere Molonglo catalog, 
and clustering of sources in the 
 2.7 GHz 2 Jy  Wall and Peacock catalog \cite{wp85} towards the supergalactic plane. 
This clustering  conforms with the association of 
UHECR arrival directions with radio sources in the supergalactic plane.

By comparison, the Auger team \cite{aug07,aug08} uses the VCV catalog \cite{vcv06}
consisting of 85,221 quasars, 1122 BL Lac objects and 21,737 
active galaxies (including 9628 Seyfert 1s).  Quoting V\'eron-Cetty and 
V\'eron, ``This [VCV] catalogue should not be used for any statistical analysis 
as it is not complete in any sense, except that it is, we hope, 
a complete survey of the literature."  The associations between 
the arrival directions of UHECRs and directions to nearby AGN in the VCV catalog may only reflect
that both the AGNs in the VCV catalog and the sources of UHECRs within $\approx 50$ -- 100 Mpc
are preferentially found toward the supergalactic plane, and that the large number of 
sources in the VCV catalog gives an apparent stronger underlying association
than would be found if a smaller catalog, such as the Wall and Peacock catalog, 
were used. Associations made with a mixed catalog are more likely to be spurious,
and the removal of statistical biases in such analyses 
are challenging. Correlating UHECR arrival directions with flux-limited
surveys of sources with relatively complete redshift information
is the only reliable way of identifying the source population(s) of UHECRs in 
the absence of high-energy neutrino detection from these sources or discovery
of anomalous $\gamma$-ray signatures in radio galaxies and GRBs with GLAST.

A concern expressed recently \cite{gor07,far08} is that although some of the UHECRs observed
with Auger are associated with the direction toward Cen A, none seem to 
originate from the Virgo cluster center of our supergalaxy in the direction of
 M87. The lower exposure of Auger toward M87, by a factor of 3 compared the exposure
to Cen A \cite{aug08}, and the Northern Hemisphere detection of UHECRs that could be associated
with Virgo \cite{sta95,cun80} 
could ameliorate this issue. 
Ref.\ \cite{far08} suggests that if all UHECRs are C, then the short energy-loss
MFP of C would mask M87 and explain preferential detection of Cen A. But our calculations
show that the effective energy loss MFP of UHECR C with $E \cong 60$ EeV is $\approx 50$
Mpc, within which M87, at a distance of $\approx 18$ Mpc, falls. 
The MFP of UHECR He  at 60 -- 100 EeV is 
$\sim 3$ -- 5 Mpc, and so could restrict arrival directions to Cen A.

The dip energy, viewed as a consequence of photopair effects on UHECR
protons \cite{ber08} either from  GRBs \cite{wda04} or AGNs \cite{alo07},  could involve 
 photodisintegration and photopair energy losses of ions interacting with photons of the EBL.
There is also the possibility, again, that 
there is a deficit
of UHECR sources between $\approx 100$ and $\approx 300$ Mpc. 
Whether this is true depends on source type, in particular whether UHECRs originate
from  specific classes of  GRBs or AGNs. 

In summary, the Auger team has opened charged particle astronomy by correlating
the anisotropic arrival directions of $\gtrsim 60$ EeV UHECRs with AGNs located within
$\approx 75$ Mpc. Calculations presented here of the effective energy-loss MFP support
an origin of UHECR ions in light nuclei with $ A \lesssim 24$.
The conclusion following from this analysis
is that detection of UHECR clustering is most easily understood if 
most UHECRs are mainly light nucleons. 
The identification
of the actual underlying source population must use complete (at least 
at high, $|b|\gtrsim 10^\circ$, galactic latitude) flux-limited catalogs.

\ack I would like to thank Justin Finke and Soebur Razzaque for many
useful conversations and for comments on the paper, 
Isabelle Grenier for sharing her presentations,
and Venya Berezinsky and the organizers of the Topics in Astroparticle 
and Underground Physics TAUP 2007 
Conference in Sendai for the opportunity to attend the conference
and visit Japan. 
This work is supported by the Office of Naval Research,  NASA
{\it GLAST} Science Investigation DPR-S-1563-Y, and  NASA {\it
Swift} Guest Investigator Grant DPR-NNG05ED411.

\bibliography{iopart-num}

\end{document}